\documentclass[journal,12pt,draftclsnofoot,onecolumn]{IEEEtran}
\usepackage{mathrsfs}
\usepackage{cite}
\usepackage{amsmath}
\usepackage{amssymb}
\usepackage{stfloats}
\usepackage{graphicx}
\usepackage{setspace}
\usepackage{xcolor}
\usepackage{multirow}
\usepackage{slashbox}

\usepackage{hhline}

\ifCLASSOPTIONcompsoc
\usepackage[tight,normalsize,sf,SF]{subfigure}
\else
\usepackage[tight,footnotesize]{subfigure}
\fi

\begin{document}
\title{Waveform Candidates for 5G Networks: Analysis and Comparison}

\author{Yinsheng~Liu, Xia Chen, Zhangdui Zhong, Bo Ai, Deshan Miao, Zhuyan Zhao, Jingyuan Sun, Yong Teng, and Hao Guan.

\thanks{Yinsheng Liu, Zhangdui Zhong, and Bo Ai are with State Key Laboratory of Rail Traffic Control and Safety, Beijing Jiaotong University, Beijing 100044, China, e-mail: \{ys.liu,zhdzhong,boai\}@bjtu.edu.cn.}
\thanks{Xia Chen is with School of Electronic and Information Engineering, Beijing Jiaotong University, Beijing 100044, China, e-mail:xchen@bjtu.edu.cn.}
\thanks{Deshan Miao, Zhuyan Zhao, Jingyuan Sun, Yong Teng and Hao Guan are with Nokia Beijing Bell Lab.}
\thanks{This research was supported by Nokia Beijing Bell Lab and natural science foundation of China under grant 61601018.}
}

\maketitle
\doublespacing

\begin{abstract}
As the next generation cellular system, 5G network is required to provide a large variety of services for different kinds of terminals, from traditional voice and data services over mobile phones to small packet transmission over massive machine-type terminals. Although orthogonal-subcarrier based waveform has been widely used nowadays in many practical systems, it can hardly meet the future requirements in the coming 5G networks. Therefore, more flexible waveforms have been proposed to address the unprecedented challenges. In this article, we will provide comprehensive analysis and comparison for the typical waveform candidates. To obtain insightful analysis, we will not only introduce the basic principles of the waveforms but also reveal the underlying characteristics of each waveform. Moreover, a comprehensive comparison in terms of different performance metrics will be also presented in this article, which provide an overall understanding of the new waveforms.
\end{abstract}


\newpage
\section{Introduction}
As the development of technology, new demands of information transmission have posed an unprecedented challenge for the next generation cellular system. In addition to the traditional voice and data services over mobile phones, 5G network is also required to support traffics that are fundamentally different from the traditional ones, such as the small packet transmission over massive machine-type communications (MMC) or Internet of things (IoT) \cite{JAndrews}.\par

As an underlying technique, a flexible waveform is required in 5G networks to address the coming challenges \cite{GWunder}. In 5G networks which will comprise many MMC scenarios, a fundamental requirement of the waveform is to support asynchronous transmission because otherwise the traffic will be filled with large overhead of synchronization signaling caused by massive terminals \cite{BFarhang}. Although orthogonal frequency division multiplexing (OFDM) has been used in long-term evolution (LTE), it can hardly meet the above requirement because the orthogonality among subcarriers cannot be maintained in asynchronous transmission. In this case, the strong out-of-band (OOB) emission of OFDM signal will cause severe adjacent-channel interference (ACI) to adjacent channels. Moreover, the synchronization signaling will also cause extra power consumption, which will reduce the lifetime of the terminals since many terminals in MMC will be driven by batteries \cite{TWild}. In addition to the asynchronous transmission, the waveform in 5G is also required to have a good localization in the time-domain in order to provide low-latency services and to support small packet transmission efficiently \cite{FSchaich}.\par

To meet the requirements of waveform in 5G networks, filter-based waveforms have been widely studied recently. The key factor that filter-based waveforms can prevail OFDM in 5G networks is that they can support asynchronous transmission by reducing the OOBs via different filter designs \cite{GWunder}. In general, the filter-based waveform can be divided into three types in terms of the filter granularity: subcarrier filtering, sub-band filtering, and full-band filtering. The subcarrier filtering based waveform, such as filtered multi-tone (FMT), is originally proposed to reduce OOB of multi-carrier signal by adopting a pair of transmit and receive filters for each subcarrier \cite{GCherubini}. Filter-bank multi-carrier (FBMC) is also a subcarrier filtering based waveform which adopts offset quadrature-amplitude-modulation (OQAM) to avoid the waste of spectrum in FMT \cite{BFarhang, BFarhang1}. By inserting a cyclic prefix (CP) in front the transmit signal, the inter-symbol interference (ISI) can be mitigated through cyclic block FMT \cite{AMTonello} and generalized frequency division multiplexing (GFDM) \cite{NMIchailow}. Despite the advantages, the long tail of the filter makes subcarrier filtering based waveforms not suitable to support low-latency service in 5G systems \cite{XWang}. Sub-band filtering based waveforms are therefore proposed to reduce the filter length. For sub-band filtering, the filters are designed with respect to a sub-band, which has more-than-one subcarriers. The length of the filter can be thus reduced since the filter bandwidth is larger than that of the subcarrier filtering. As a sub-band filtering based waveform, the sub-band of resource block filtered OFDM (RB-F-OFDM) is as wide as a RB, which contain $12$ subcarriers in LTE \cite{JLi,JLi1}. Universal filtered multi-carrier (UFMC) is another type of sub-band filtering based waveform, where only a transmit filter is used while the demodulation in the receiver relies on the oversampled discrete Fourier transform (DFT) \cite{XWang,FSchaich,TWild}. The filters can be also designed with respect to the full bandwidth such that only one filter is enough as in filtered OFDM (F-OFDM). In this case, F-OFDM will be equivalent to the RB-F-OFDM if only one RB is available in the bandwidth \cite{JAbdoli, XZhang}.\par

In this article, we will make a comprehensive comparison of different types of waveforms. Since many new waveforms have been proposed and we cannot take everything into account in one article, we only choose four typical waveform candidates, that is, FBMC, RB-F-OFDM, UFMC, and F-OFDM, because they are not only typical representatives but have also gained more attention from the industry and the academia. Our comparisons in this article are based on insightful analysis of different waveforms. We will not only introduce the basic principles but also reveal the underlying characteristics of each waveform.\par

The rest of this article is organized as follows. In Section II, we will introduce the basic principle of various waveform candidates. In Section III, we will introduce the applications of those waveforms in wireless channels. A comprehensive comparison will be presented in Section IV, and finally summarization and conclusions are in Section V.
\begin{figure}
  \centering
  \includegraphics[width=5.7in]{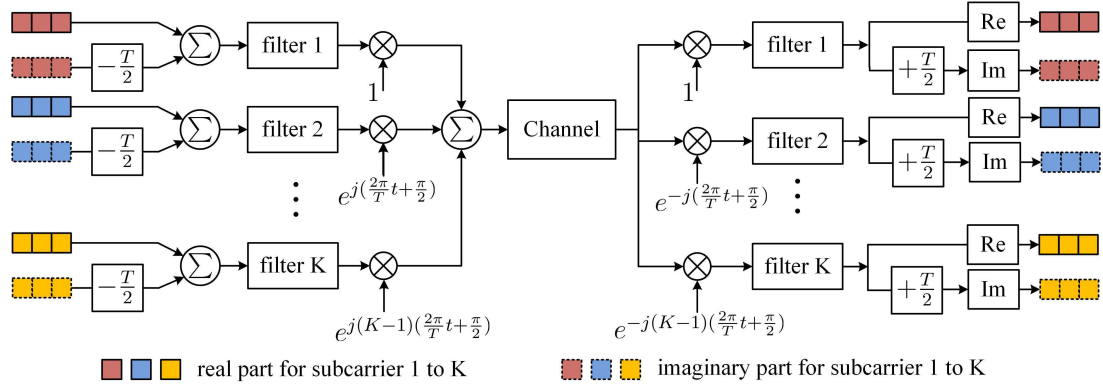}\\
  (a) FBMC\\
  ~\\
  \includegraphics[width=5.7in]{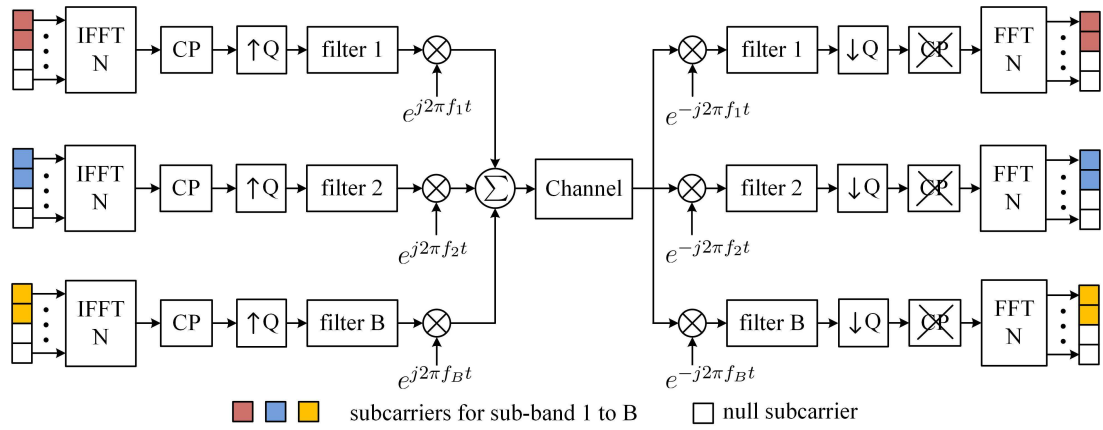}\\
  (b) RB-F-OFDM\\
  ~\\
  \includegraphics[width=4.6in]{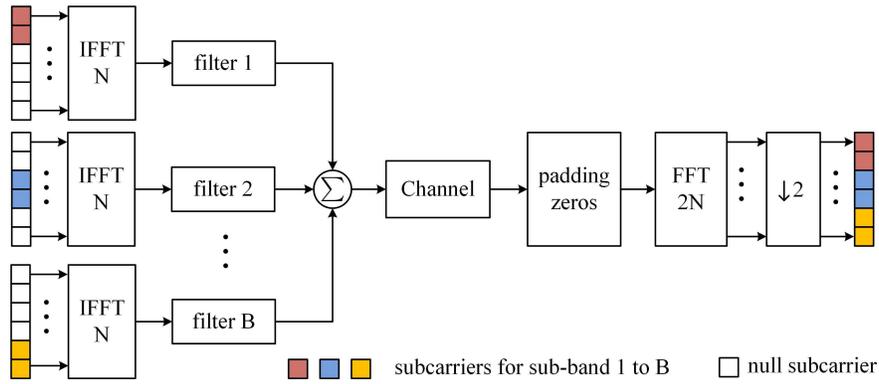}\\
  (c) UFMC\\
  ~\\
  \includegraphics[width=4.1in]{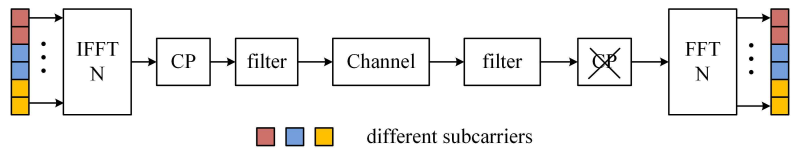}\\
  (d) F-OFDM
  \caption{System structures for (a) FBMC, (b) RB-F-OFDM, (c) UFMC, and (d) F-OFDM.}\label{fig1}
\end{figure}
\section{Principles of Waveform Candidates}
In this section, we will introduce the basic principles of FBMC, RB-F-OFDM, UFMC, and F-OFDM, respectively. Since different users can be separated by the filters, we only consider the single-user case in this article.

\subsection{Subcarrier filtering: FBMC}
The system structure for FBMC is shown in Fig.~\ref{fig1} (a), where $T$ denotes the time interval of FBMC symbols and $K$ denotes the number of subcarriers. For each sbucarrier, the complex QAM symbols are first split into real parts and imaginary parts. After a time delay of $T/2$ for the imaginary parts, the combined OQAM signals are fed to the transmit filter, and then modulated by the corresponding subcarrier frequency before sending to the channel. The receiver follows a reverse procedure for signal demodulation.\par

\begin{figure}
  \centering
  \includegraphics[width=5.2in]{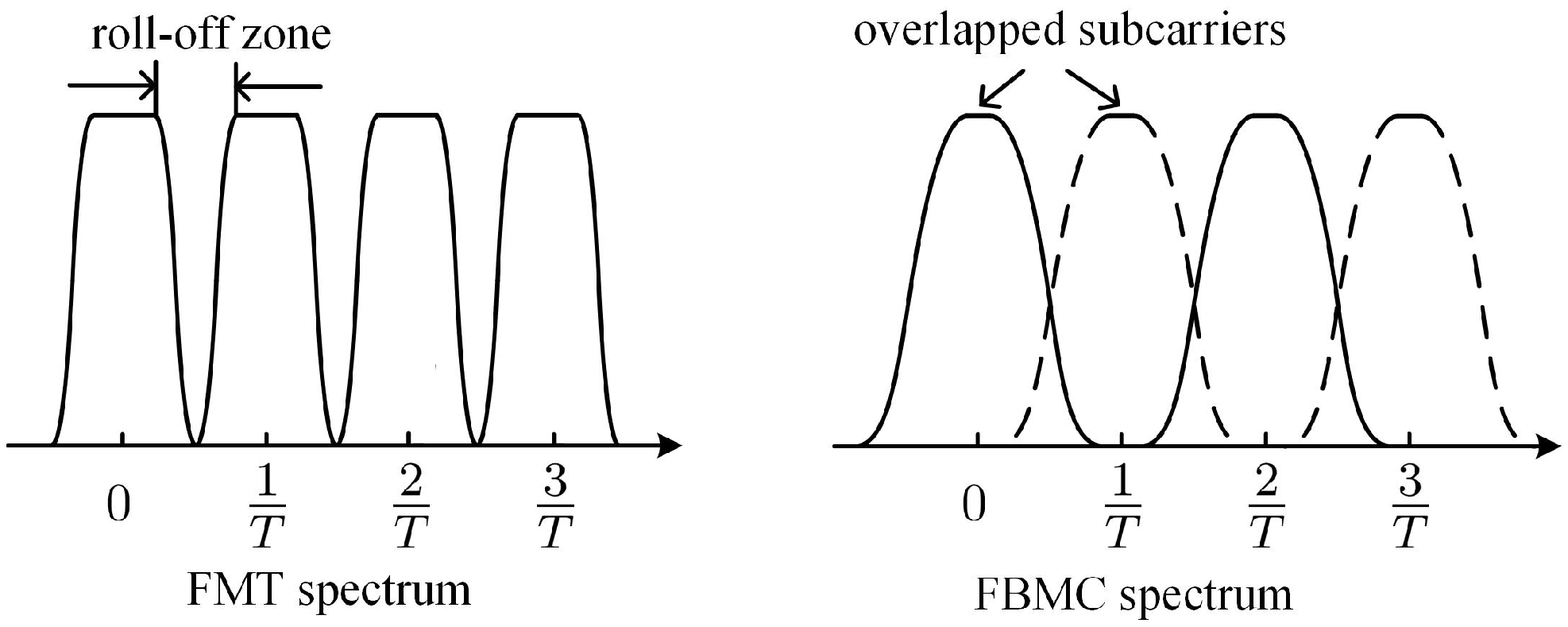}\\
  ~~~~~~~~~~~~~~~~~~~~~~~~~~~~~~~~~~~~~~~~~~~~(a)~~~~~~~~~~~~~~~~~~~~~~~~~~~~~~~~~~~~~~~~~~~~\\
  \includegraphics[width=5in]{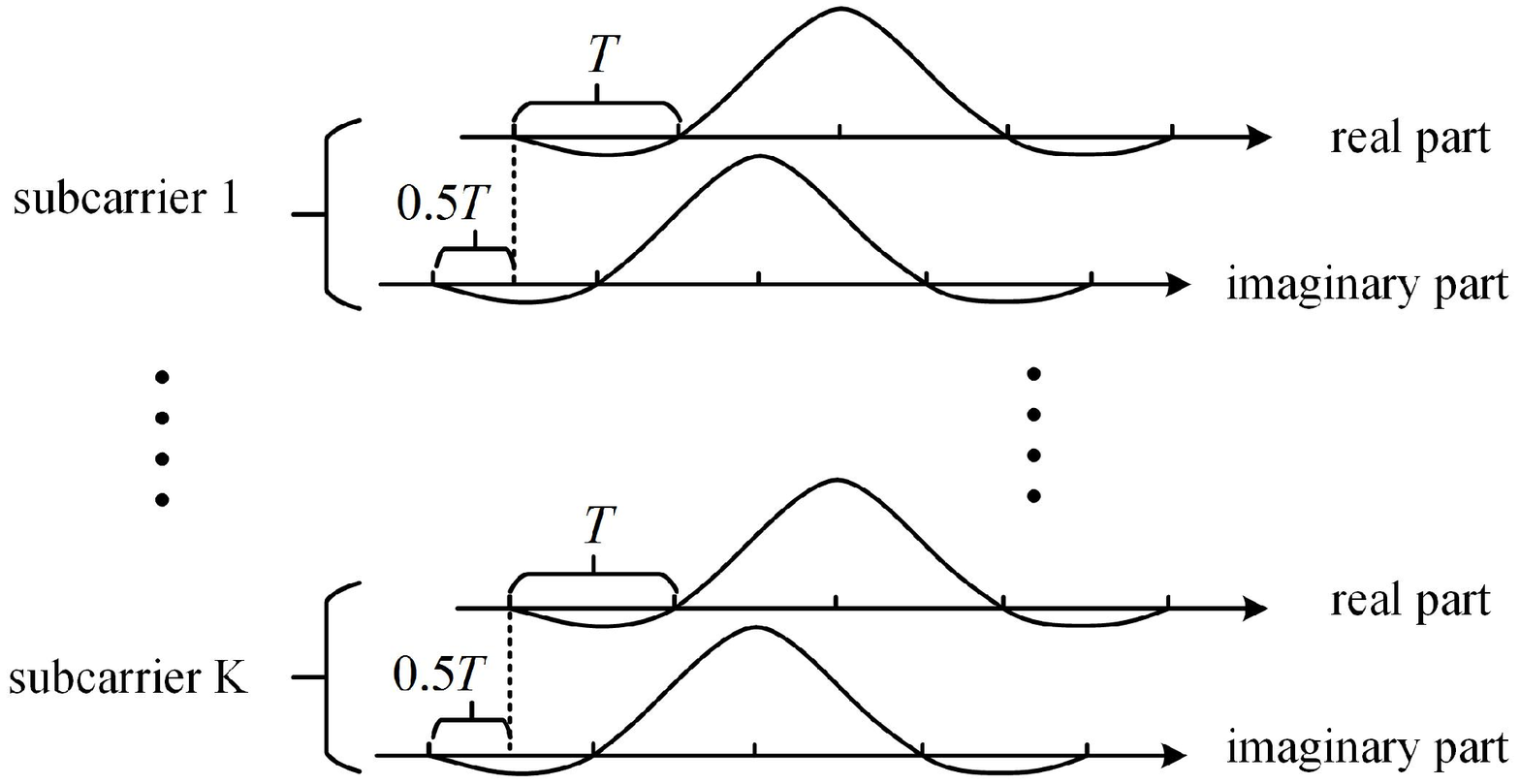}\\
  ~~~~~~~~~~~~~~~~~~~~~~~~~~~~~~~~~~~~~~~~~~~~(b)~~~~~~~~~~~~~~~~~~~~~~~~~~~~~~~~~~~~~~~~~~~~
  \caption{Waveform property for FBMC with (a) spectrums for FMT and FBMC and (b) impulse response of the transmit (receive) filter.}\label{spectrum}
\end{figure}

For a general multi-carrier system such as FMT, the orthogonality among subcarriers cannot hold due to the transmit and the receive filters. To avoid the ICI, the filter spectrums for different subcarriers should have no overlap \cite{GCherubini}. This will result in a waste of the spectrum since roll-off transitions are required in practical raised cosine filters, as in Fig.~\ref{spectrum} (a). To make full use of the available spectrum, FBMC first relaxes the requirement of the filter design by allowing spectrum overlapping for adjacent subcarriers, as in Fig.~\ref{spectrum} (a). Then, the OQAM scheme in FBMC makes sure that the interference caused by adjacent subcarriers can be removed by recovering the real and the imaginary parts separately. In this manner, FBMC can achieve the same spectrum efficiency as the standard OFDM.\par

Due to the subcarrier filtering, the tail of the filter impulse response in FBMC systems will typically cover four symbol intervals, which is much longer than the other filter-based waveforms. Compared to other waveforms, the long tail makes FBMC not suitable for short small packet transmission and low latency service. The data transmission and reception can happen immediately within a symbol interval through standard OFDM or other filter-based waveforms. For FBMC as shown in Fig.~\ref{spectrum} (b), however, the data transmission will be postponed due to the latency caused by the long tail of the filter response, and the reception cannot be finished until the whole pulse has been received. In addition, due to the long tail, the demodulation of FBMC symbols relies on the channel response over multiple symbol intervals. As a result, the time variation of the channel responses will be averaged out, making it more difficult to gain Doppler diversity in FBMC systems. We will further elaborate this issue through simulation results in Section IV.

\subsection{Sub-band filtering: RB-F-OFDM and UFMC}
The sub-band filtering based waveform is proposed to overcome the drawbacks of the subcarrier filtering. Since the filters are designed with respect to multiple subcarriers, the length of the filter impulse response is reduced compared to the subcarrier filtering based one. Therefore, sub-band filtering based waveform can be used to serve low latency applications in 5G network.\par

\subsubsection{RB-F-OFDM}
The system structure of RB-F-OFDM is shown in Fig.~\ref{fig1} (b) where $N$ denotes the size of inverse fast Fourier transform (IFFT)/FFT. The data symbols for different subcarriers at each sub-band are first converted to time domain through a standard OFDM modulation. After upsampling, the generated signal is sent to the transmit filter and then shifted to the carrier frequency of the corresponding sub-band. The receiver follows a reverse procedure for signal demodulation. The upsampling and corresponding downsampling are used to reduce the sampling rate of FFT/IFFT. Accordingly, the sizes of IFFT/FFT are also reduced, leading to a low implementation complexity.\par

Similar to the subcarrier filtering, when multiple RBs are adjacent in the frequency domain, spectrum leakages from other RBs cannot be avoided due to the transition zone of the transmit and receiver filters. Even though, the spectrum leakage will be very small due to the transmit and receive filtering. Therefore, the inter-user interference caused by the spectrum leakage will be very small and thus can be omitted.\par

\subsubsection{UFMC}

\begin{figure}
  \centering
  \includegraphics[width=5in]{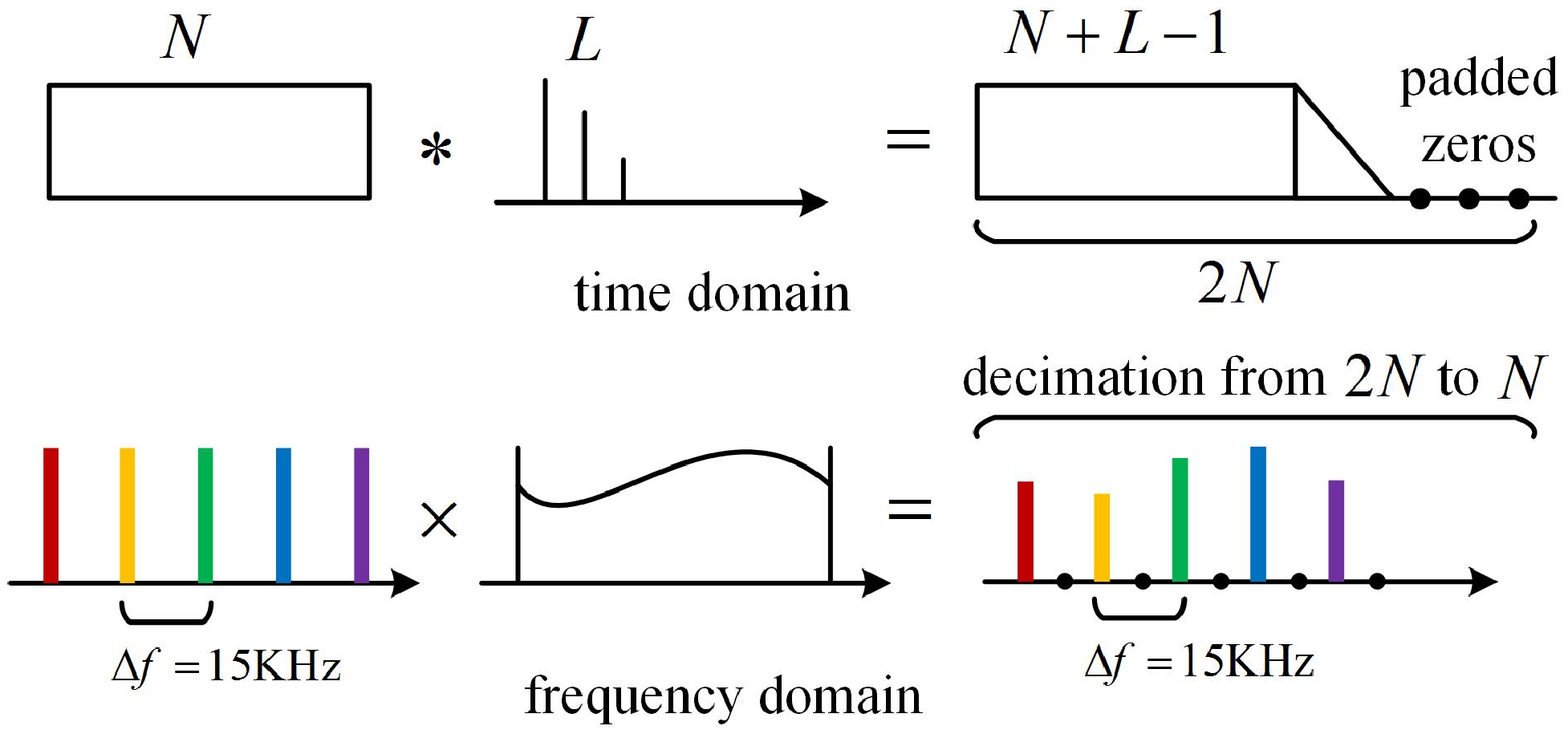}\\
  ~~~~~~~~~~~~~~~~~~~~~~~~~~~~~~~~~~~~~~~~~~~~(a)~~~~~~~~~~~~~~~~~~~~~~~~~~~~~~~~~~~~~~~~~~~~\\
  \includegraphics[width=5.5in]{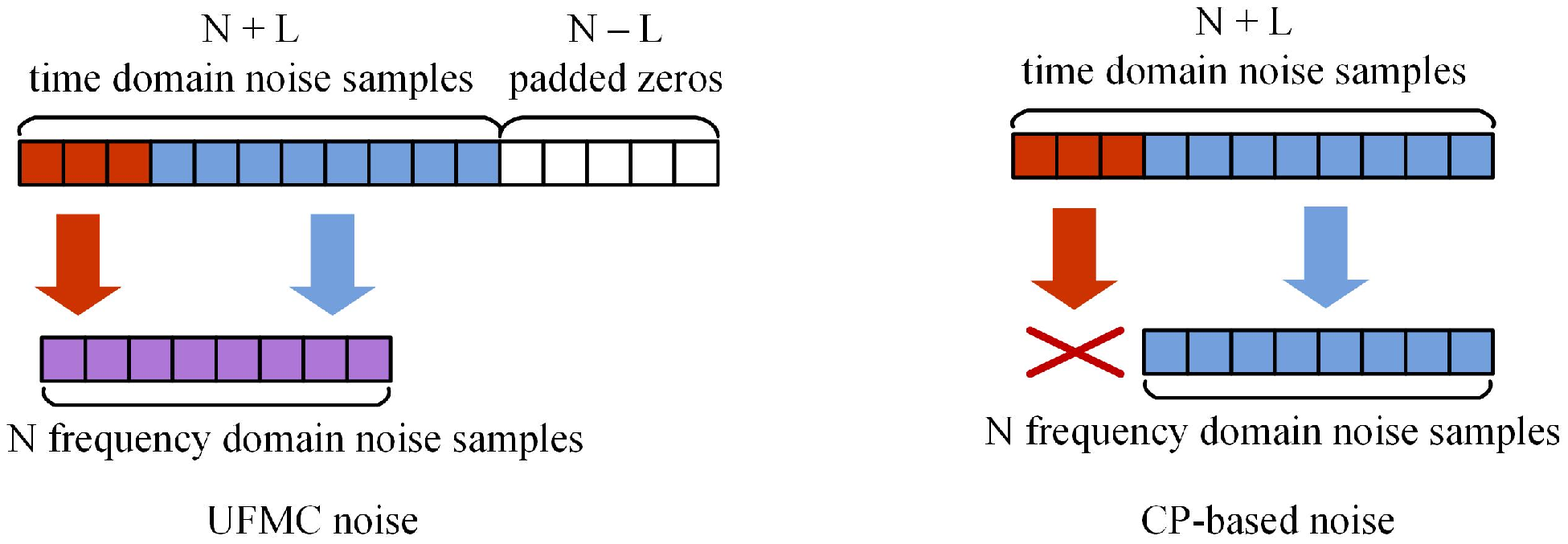}\\
  ~~~~~~~~~~~~~~~~~~~~~~~~~~~~~~~~~~~~~~~~~~~~(b)~~~~~~~~~~~~~~~~~~~~~~~~~~~~~~~~~~~~~~~~~~~~
  \caption{UFMC property with (a) $2N$-FFT based reception and (b) explanation of noise enhancement.}\label{OverFFT}
\end{figure}

UFMC is another type of sub-band filtering based waveform. As shown in Fig.~\ref{fig1} (c) where $N$ is the size of the IFFT at the transmitter, the transmitter has a similar structure with that in RB-F-OFDM, except the insertion of CP and the upsampling. Accordingly, the frequency shift is also not required because the data symbols have been mapped to the corresponding sub-bands before IFFT. Different from other filter-based waveforms where receive filters are also employed, UFMC only uses the transmit filters and the signal demodulation depends on a $2N$-point FFT. A key feature of UFMC is the employment of the $2N$-point FFT for signal reception, which can demodulate the data symbols without the need of CP. Given the length of the transmitted signal, $N$, and the length of the channel impulse, $L$, the received signal will be $N+L-1$. By padding $N-L-1$ zeros at the end of the received signal, the data symbols can be recovered through the $2N$-point FFT followed by a decimation. Fig.~\ref{OverFFT} (a) illustrates the procedure of the $2N$-point FFT based demodulation.

Although $2N$-point FFT can achieve efficient demodulation without the need of CP, it also causes a noise enhancement to the UFMC reception compared to other CP-based waveforms. As an explanation, we consider an additive white Gaussian noise (AWGN) channel for simplicity. Assume that the filter length is $L$ which is equal to the CP length, then $N+L$ time-domain noise samples contribute to the frequency domain noise, as in Fig.~\ref{OverFFT} (b). On the other hand, since the CP is removed for CP-based waveform before further processing, the additive noise included in the CP are also removed and thus only $N$ time-domain noise samples contribute to the frequency domain noise. As a result, the noise power in UFMC will be larger than the other CP-based waveforms. Due to the noise enhancement, the signal-to-noise ratio (SNR) of UFMC will be degraded by $10\lg\left(1+L/N\right)~(\text{dB})$, which corresponds to $0.33$ dB for normal CP length and $0.97$ dB for extended CP in current LTE standard.
\subsection{Full-band filtering: F-OFDM}
For full-band filtering, only one filter is required even though multiple RBs can be allocated to the user of interest. As shown in Fig.~\ref{fig1} (d), F-OFDM can be viewed as an extension of the standard OFDM by adopting a pair of transmit and receiver filters at the transmitter and the receiver respectively.\par

When multiple RBs are allocated to the user of interest, the OOB depends on the pattern of the RBs. If the allocated RBs are adjacent in the frequency domain, they comprise a single but wider frequency block and thus the OOB can be effectively reduced through the transmit filter. When the RBs are non-adjacent, however, the OOB between the RBs is in the pass band of the transmit filter and thus it cannot be reduced effectively \cite{JAbdoli}. On the other hand, if only one RB is available in the bandwidth, Fig.~\ref{fig1} shows that F-OFDM will have a similar structure with RB-F-OFDM except that upsampling and decimation are used in RB-F-OFDM to reduce implementation complexity. In this case, the OOB can be always reduced.

\section{Application in Wireless Channels}
The principles of different waveforms in the above section are introduced under the assumption of AWGN channel. In this section, we will analyze the impact of ISI caused by multipath propagation in practical wireless channels, and the corresponding channel estimation and equalization techniques.

\subsection{ISI Analysis}
In standard OFDM, CP is used to protect the ISI from the adjacent OFDM symbols. Usually, the length of CP should be larger than or equal to the maximum delay spread such that the ISI can be completely removed. Even though the CP or the guard interval are also used in filter-based waveforms, the length of the effective channel impulse response (CIR), which is composed of the transmit and receive filters and the real channel, is usually larger than the length of the CP or the guard interval. As a result, ISI cannot be avoided in filter-based waveforms. It is therefore necessary to have an analysis on the impact of ISI.\par

For simplicity, we assume an OFDM transmission with $N$-point FFT and $K$ subcarriers are used for data transmission $(K<N)$. Without loss of generality, we assume that the ISI is caused by $L$ taps of the effective CIR with unit total power. Then, the signal-to-interference plus noise (SINR) can be approximated by $\mathrm{SINR}=\left(LKN^{-2}+\mathrm{SNR}^{-1}\right)^{-1}$. In practical systems, we always have $N^2\gg LK$ and therefore $\mathrm{SINR}\approx \mathrm{SNR}$. In fact, the symbol duration in multi-carrier systems are usually much larger than the maximum delay spread. Therefore, even though the ISI cannot be removed, its impact is very small and thus can be ignored. The above analysis can be also used for the other filter-based waveforms.

\subsection{Channel Equalization}
The key advantage of a multi-carrier system is that it can greatly simplify the tasks of channel estimation and equalization. From above analysis, even though the ISI cannot be removed perfectly, it can be still omitted in filter-based waveforms. As a result, like standard OFDM, the channel estimation and equalization for filter-based waveforms can be also conducted with a single tap equalizer for each subcarrier in the frequency domain.\par

For full-band filtering and sub-band filtering, the schemes of the waveforms are similar to the standard OFDM. Therefore, traditional channel estimation and equalization techniques, which have been widely addressed in existing works, can be directly used in such systems. For FBMC, channel estimation and equalization can be also conducted with respect to each subcarrier. However, due to the restriction on the FBMC modulation, channel estimation and equalization in FBMC systems are less explored than in traditional OFDM systems \cite{BFarhang}.

\section{Comparison}
For insightful understanding of different waveforms, a comprehensive comparison is presented in this section. Numerical simulations are conducted to assess the OOB and the block error rate (BLER). For the simulation, we assume the carrier frequency is $2$ GHz. The subcarrier spacing is $15$ KHz with $15.36$ MHz sampling frequency (corresponds to $1024$ FFT in current LTE standard), and $36$ subcarriers (corresponds to $3$ RBs) are employed for data transmission. In particular, the sub-band of UFMC is as wide as one RB for easy comparison. A normal CP length is used when CP is required. Quadrature-phase-shift-keying (QPSK) and 16QAM are both considered in the simulation with $1/3$ rate turbo coding. For channel modeling, the extended typical urban (ETU) channel model is used with a maximum delay spread of $5\mu$s. To focus on the link-level performance, we only consider a single user in one cell with single transmit antenna at the user terminal and single receive antenna at the base station. Perfect channel estimation are used in the simulation so that we can highlight the performance difference caused by the schemes of the waveforms.\par

\subsection{OOB}
\begin{figure}
  \centering
  \includegraphics[width=3.2in]{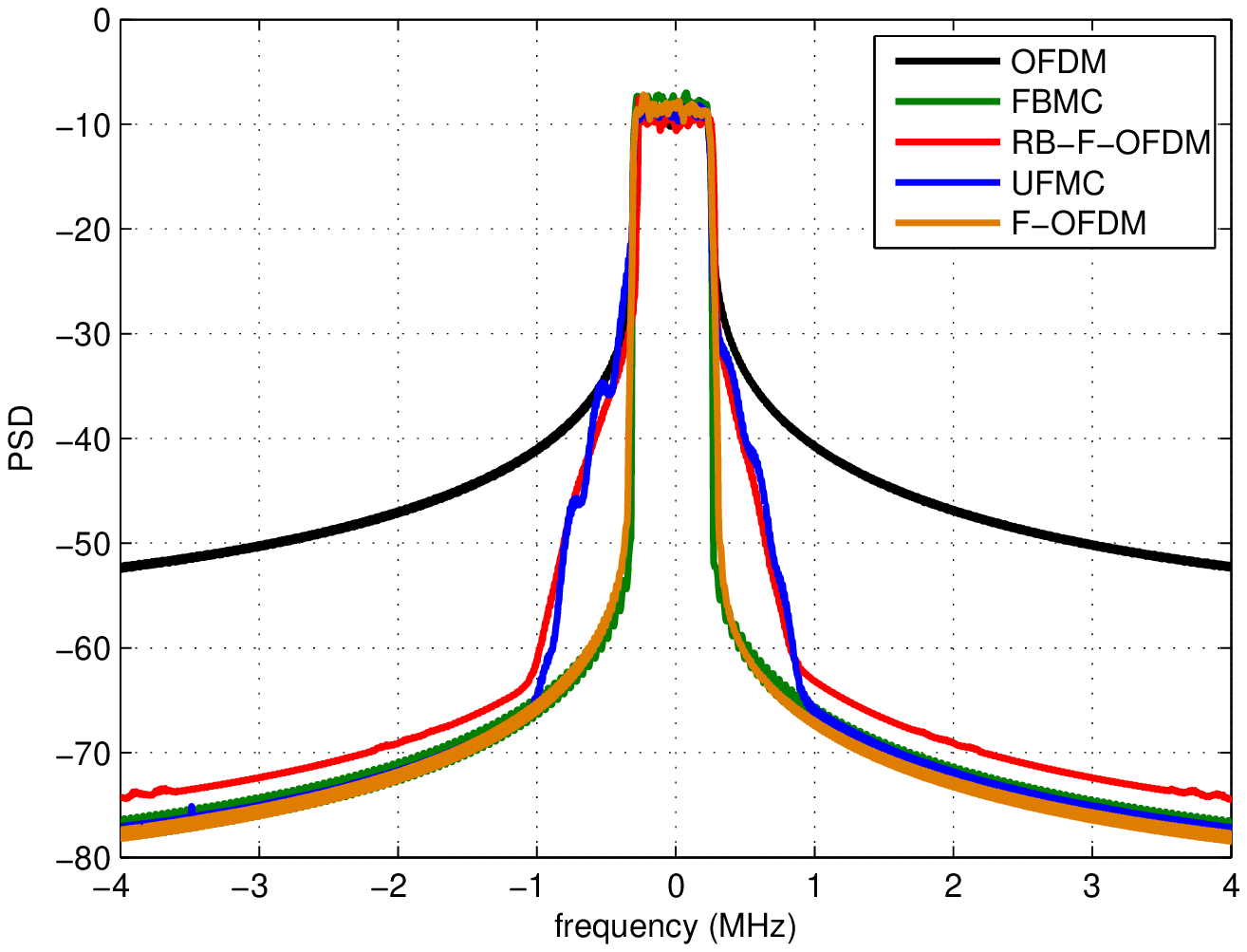}
  \includegraphics[width=3.2in]{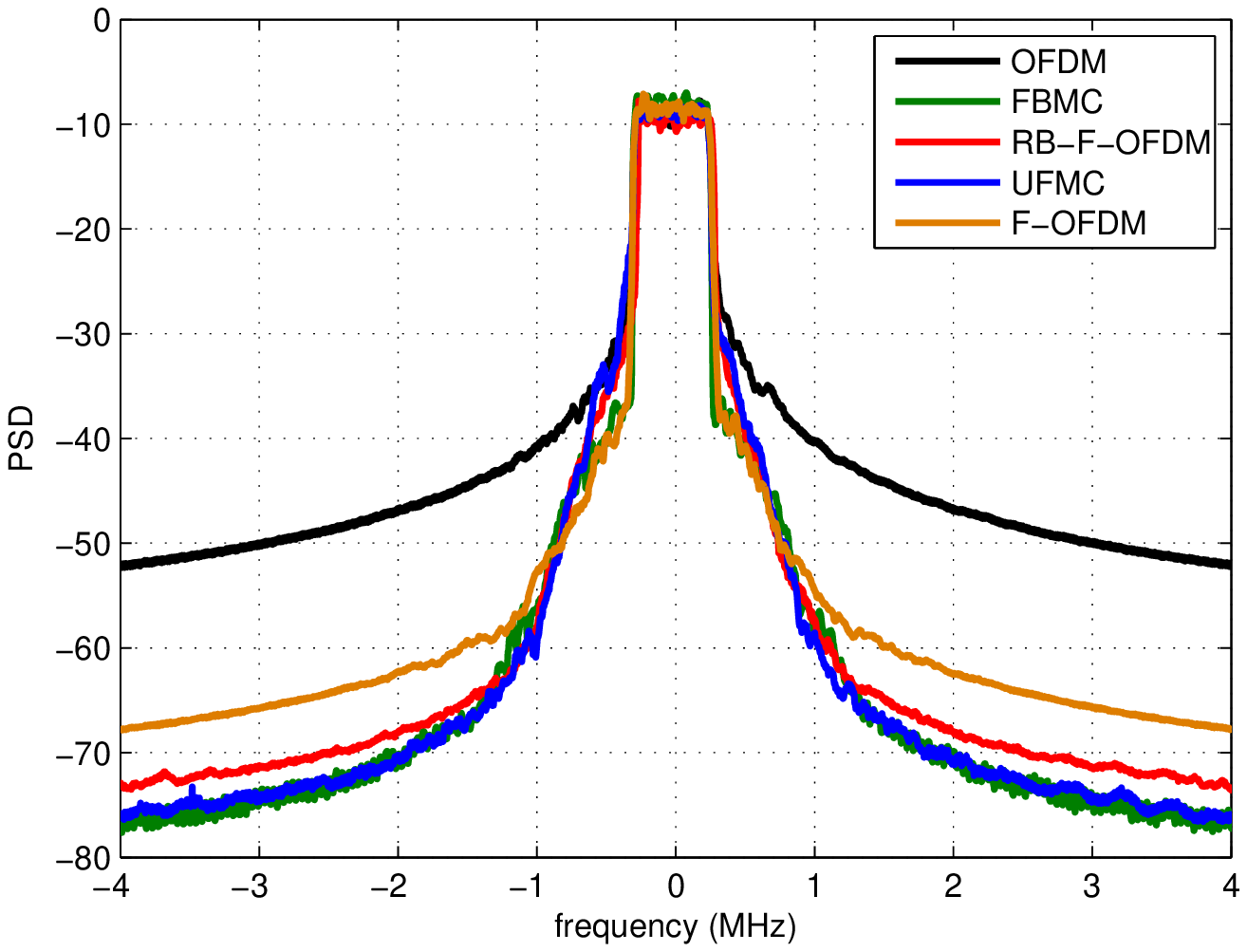}\\
  (a)~~~~~~~~~~~~~~~~~~~~~~~~~~~~~~~~~~~~~~~~~~~~~~~~~~~~~~(b)\\
  \includegraphics[width=3.2in]{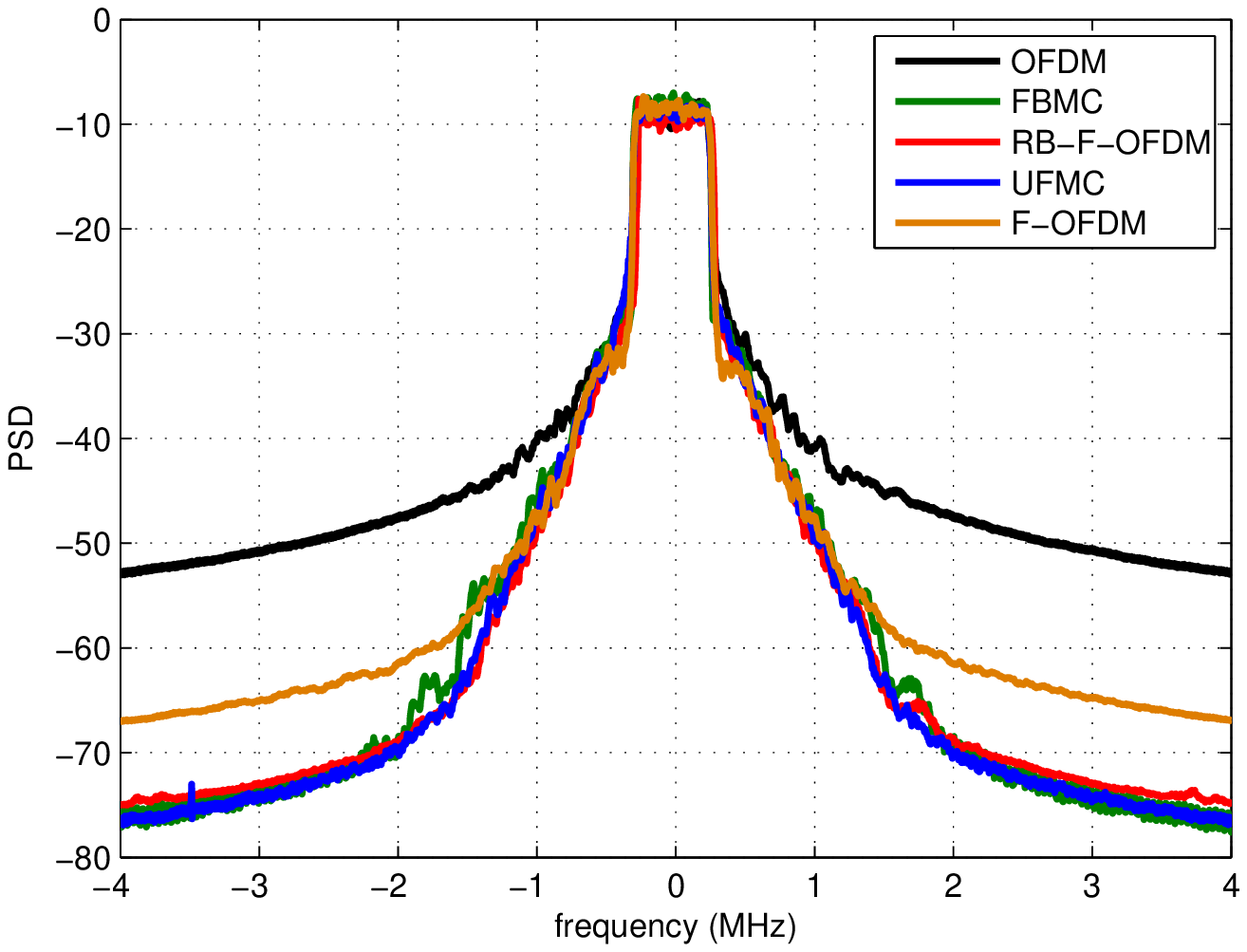}
  \includegraphics[width=3.2in]{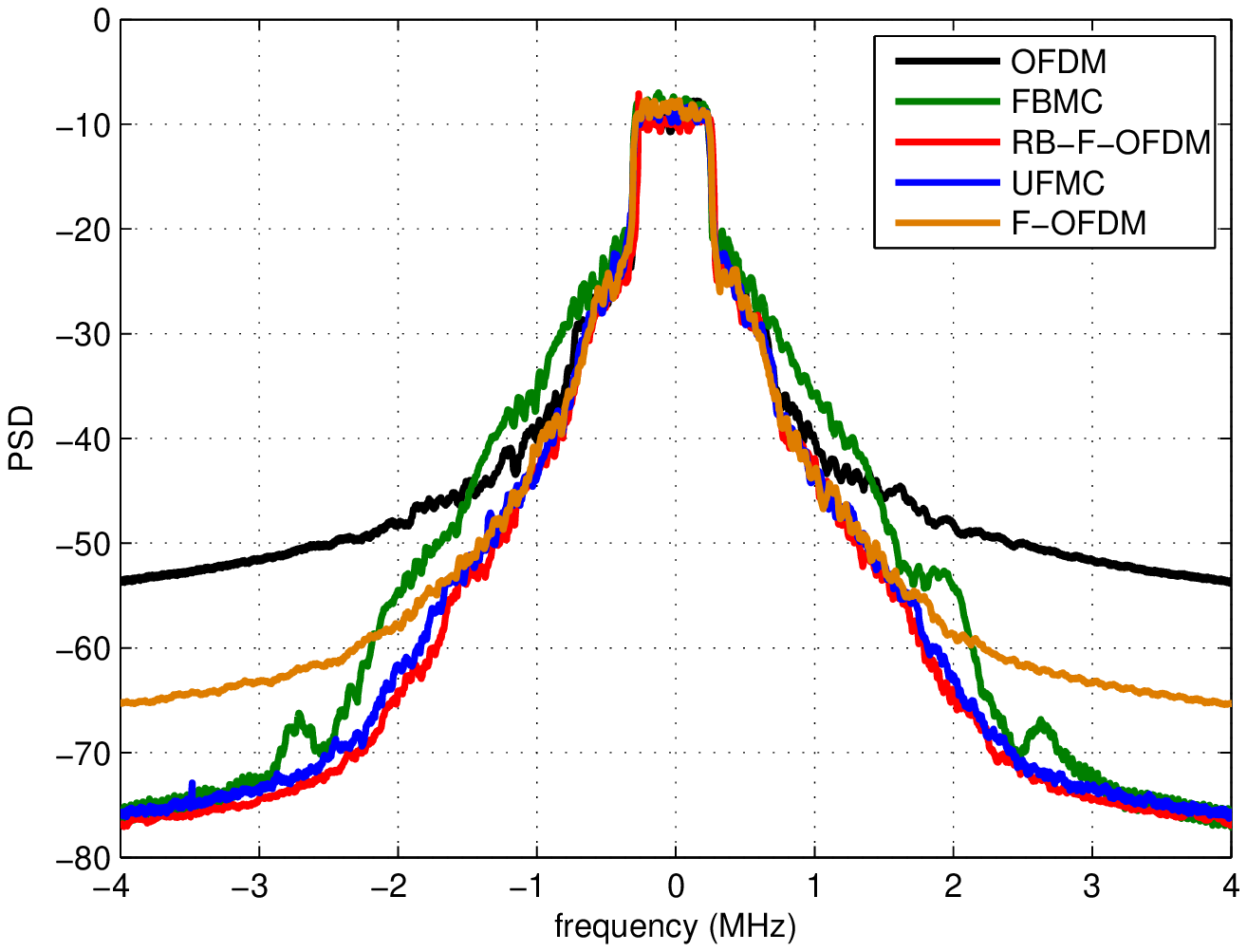}\\
  (c)~~~~~~~~~~~~~~~~~~~~~~~~~~~~~~~~~~~~~~~~~~~~~~~~~~~~~~(d)
  \caption{PSDs for different waveforms under different conditions (a) no PA (b) $20$ dBm output power with PA (c) $25$ dBm output power with PA (b) $29$ dBm output power with PA. The PSDs have been normalized so as to be presented with the same scale.}\label{PSD}
\end{figure}
The key advantage of filter-based waveforms is that they can significantly reduce the OOB to support asynchronous transmission. It is therefore necessary to compare the OOBs of different waveforms with the standard OFDM.\par

Fig.~\ref{PSD} shows the power spectrum density (PSD) for different waveforms. The nonlinearity of the power amplifier (PA) is also taken into account \cite{PA}. From Fig.~\ref{PSD} (a) where ideal linear PA is assumed, the OOBs of filter-based waveforms can achieve $15-20$ dB improvement over the standard OFDM. In this case, the OOB of the waveform depends heavily on the filter design. On the other hand, however, if the nonlinear of PA is taken into account, the OOB will be mainly determined by the output power and thus the OOB of different waveforms will be similar, as shown from Fig.~\ref{PSD} (b) to (c). The nonlinearity of the PA will cause extra frequency components and thus the OOBs will increase accordingly. It is also shown that the OOB will be worse as the rising of the output power because the PA with larger output power will be closer to the nonlinear zone.\par

Our results show that all kinds of filter-based waveforms can significantly reduce the OOB compared to the standard OFDM, even in the presence of the nonlinearity of PA. In this sense, the filter-based waveforms are more suitable for asynchronous transmission because the ACI can be greatly reduced.

\subsection{BLER}

As stated from above section, the employment of transmit filter or receive filter will increase the length of the effective CIR, leading to extra ISI to the filer-based waveforms. In this subsection, BLERs of different waveforms are assessed with respect to SNR, carrier-frequency-offset (CFO), and terminal mobility. In practical systems, the performance depends heavily on the nonlinearity of the PA. However, in order to highlight the difference caused by the schemes of various waveforms, the ideal linear PA is considered in the simulation to avoid the impact caused by the PA nonlinearity.

\subsubsection{BLER versus SNR}
\begin{figure}
  \centering
  \includegraphics[width=5.5in]{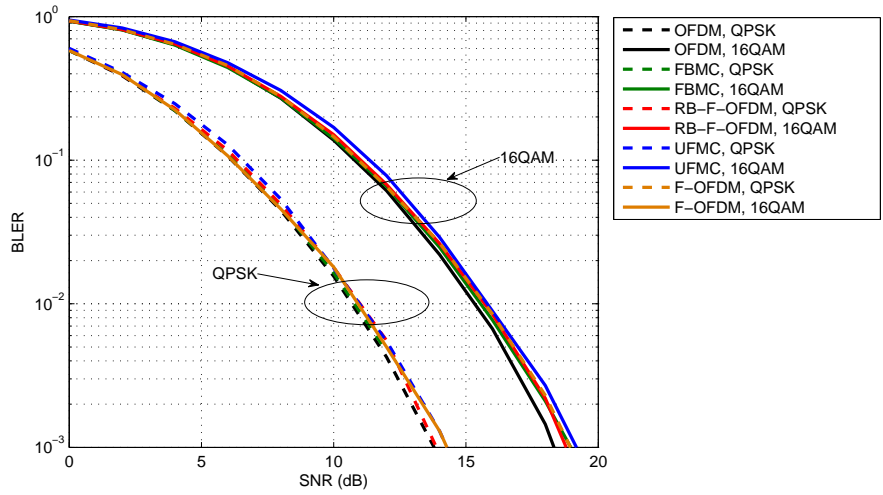}\\
  ~~~~~~~~~~~~~~~~~~~~~~~~~~~~~~~~~~~~~~~~~~~~(a)~~~~~~~~~~~~~~~~~~~~~~~~~~~~~~~~~~~~~~~~~~~~\\
  \includegraphics[width=5.5in]{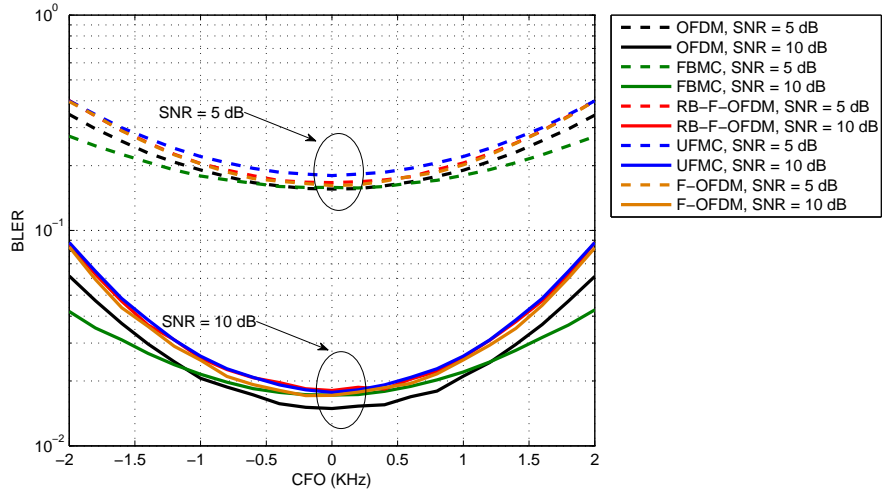}\\
  ~~~~~~~~~~~~~~~~~~~~~~~~~~~~~~~~~~~~~~~~~~~~(b)~~~~~~~~~~~~~~~~~~~~~~~~~~~~~~~~~~~~~~~~~~~~\\
  \includegraphics[width=5.5in]{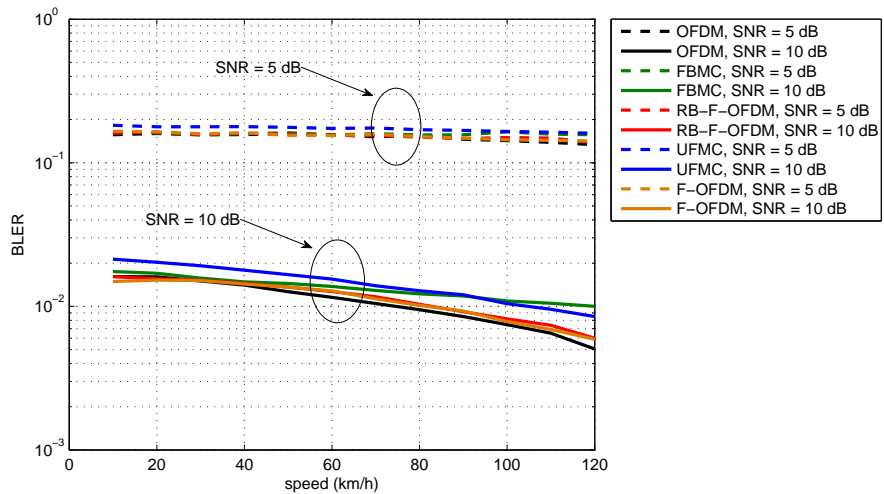}\\
  ~~~~~~~~~~~~~~~~~~~~~~~~~~~~~~~~~~~~~~~~~~~~(c)~~~~~~~~~~~~~~~~~~~~~~~~~~~~~~~~~~~~~~~~~~~~\\
  \caption{BLERs for different waveforms versus (a) SNR (b) CFO and (c) terminal speed.}\label{BLER}
\end{figure}

Fig.~\ref{BLER} shows the block-error rate (BLER) versus SNRs with different modulation schemes. As expected, standard OFDM can achieve the best performance since ISI has been removed due to CP. For filter-based waveforms, however, performance degradation can be observed because they will suffer from extra ISI. This is especially the case for high-order modulation because the dense constellation makes it more sensitive to the interference. Even though, the performance degradation is very tiny because the ISI is very small, as we have analyzed in Section III. We can also observe that the UFMC has the worst performance among all the waveforms because of the noise enhancement caused by $2N$-point FFT based reception. Our results in Fig.~\ref{BLER} show that performance degradation due to noise enhancement is about $0.3$ dB, which coincides with our theoretical analysis in Section III.

\subsubsection{CFO Robustness}

In multi-carrier systems, CFO will cause severe inter-carrier interference (ICI) and thus degrade the performance. It is therefore necessary to evaluate the robustness of different waveforms in the presence of CFO. Fig.~\ref{BLER} (b) shows the BLER versus CFO for different waveforms with different SNRs. In general, the performance will degrade as the increasing of CFO. This is especially the case for the high SNR case because the additive noise is small in this case and thus the interference will be dominant. From the figure, FBMC is the most robust against the CFO. This is because the transmit filter of FBMC has a much better frequency-domain localization than the other waveforms \cite{BFarhang1}, and therefore the ICI can be filtered out efficiently for each subcarrier, leading to much smaller ICI. Theoretically, the ICI can be also reduced for sub-band filtering based waveforms because the interference from other sub-bands can be removed through receive filters. In practice, however, Fig.~\ref{BLER} (b) shows that the sub-band filtering based waveform suffers from similar interference with OFDM. This is because the ICI is mainly determined by the adjacent subcarriers that are in the same sub-band with the subcarrier of interest and thus cannot be effectively removed through sub-band filtering. For F-OFDM, the ICI will be the same with the standard OFDM since the full-band filtering cannot remove any ICI.

\subsubsection{Mobility}

Support for high mobility users is an important feature of the future 5G system. It is therefore necessary to investigate the impact of terminal mobility on different waveforms. Fig.~\ref{BLER} (c) evaluates the performance with respect to different terminal speeds up to $120$ km/h, which corresponds to a maximum Doppler frequency of about $222$ Hz for a $2$ GHz carrier frequency. Since we have assumed perfect channel estimation in the simulation, the performance can be improved by increasing the terminal speed because more Doppler diversity can be captured by the receiver for higher speed. This is especially the case for the high SNR situation because the diversity gain will be more obvious when the noise power is small. However, compared to other waveforms, the performance of FBMC only improves a little even when the terminal speed is high to $120$ km/h. This is because the time variation of channels over adjacent symbol intervals have been averaged out due to the long tail of filter impulse response. It is therefore more difficult for FBMC to gain the Doppler diversity.

The BLER results show that even though the employment of the transmit or receive filter will cause extra ISI to the filter-based waveforms, the performance degradation is actually very small and thus can be neglected.

\subsection{Support for Low-latency Service}
In addition to the low OOB, the good time localization is also required for 5G waveforms to support low-latency services. From the above section, the schemes of RB-F-OFDM, UFMC, and F-OFDM are actually similar to that of standard OFDM, and thus the data transmission can happen immediately within a symbol interval. However, the data transmission for FBMC will be postponed due to the latency caused by the long tail of the filter response, and the reception cannot be finished until the whole pulse has been received. In this sense, FBMC is not as suitable as the other filter-based waveforms for low-latency services.

\begin{table}
  \centering
  \begin{tabular}{|c||c|c|c|c|c|}
     \hline
     \multirow{2}{*}{\backslashbox{Metrics}{Waveform}} & \multirow{2}{*}{OFDM} & \multirow{2}{*}{FBMC} & \multirow{2}{*}{RB-F-OFDM} & \multirow{2}{*}{UFMC} & \multirow{2}{*}{F-OFDM}\\
     & & & & & \\
     \hhline{|=||=|=|=|=|=|}
     \multirow{2}{*}{Filter granularity} & \multirow{2}{*}{$-$} & \multirow{2}{*}{subcarrier} & \multirow{2}{*}{sub-band} & \multirow{2}{*}{sub-band} & \multirow{2}{*}{full-band}\\
      & & & & & \\
     \hline
     \multirow{2}{*}{OOB without PA} & \multirow{2}{*}{C} & \multirow{2}{*}{A} & \multirow{2}{*}{B} & \multirow{2}{*}{B} & \multirow{2}{*}{A}\\
      & & & & & \\
      \hline
     \multirow{2}{*}{OOB with PA} & \multirow{2}{*}{C} & \multirow{2}{*}{B} & \multirow{2}{*}{B} & \multirow{2}{*}{B} & \multirow{2}{*}{B}\\
      & & & & & \\
      \hline
     \multirow{2}{*}{Low latency} & \multirow{2}{*}{A} & \multirow{2}{*}{C} & \multirow{2}{*}{A} & \multirow{2}{*}{A} & \multirow{2}{*}{A}\\
     & & & & & \\
     \hline
     \multirow{2}{*}{BLER vs SNR} & \multirow{2}{*}{A} & \multirow{2}{*}{B} & \multirow{2}{*}{B} & \multirow{2}{*}{C} & \multirow{2}{*}{B}\\
      & & & & & \\
     \hline
     \multirow{2}{*}{CFO robustness} & \multirow{2}{*}{B} & \multirow{2}{*}{A} & \multirow{2}{*}{B} & \multirow{2}{*}{B} & \multirow{2}{*}{B}\\
     & & & & & \\
     \hline
     \multirow{2}{*}{Doppler diversity} & \multirow{2}{*}{A} & \multirow{2}{*}{B} & \multirow{2}{*}{A} & \multirow{2}{*}{A} & \multirow{2}{*}{A} \\
     & & & & & \\
     \hline
   \end{tabular}
  \caption{Summarization for different waveforms. Grades A to C indicate from good to bad.}\label{Sum}
\end{table}

\section{Summarization and Conclusions}

As a summarization, different features of various waveforms have been listed in Tab.~\ref{Sum} where grades A to C indicate from good to bad. To achieve asynchronous uplink transmission, all waveforms rely on the transmit and receiver filters with different filter granularity. All kinds of filter-based waveform can achieve lower OOBs than the standard OFDM even in the presence of PA nonlinearity. Due to the extra ISI, the filter-based waveforms will suffer from small performance degradations compared to that of standard OFDM. In particular, UFMC has the worst performance because it has to suffer from an extra noise enhancement. Although the per-subcarrier filtering feature makes FBMC most robust against CFO, it also makes FBMC hard to obtain the Doppler diversity since the channel time variation has been averaged out by the long tail of filter impulse. Meanwhile, the long tail of the filter also makes FBMC not suitable for low-latency services.\par

Due to the requirement of asynchronous transmission in future 5G networks, filter-based waveforms have gained a lot attention recently. In this article, we have discussed the basic principles and revealed the underlying characteristics of typical filter-based waveforms. In addition, a comprehensive analysis has also been presented in this article to provide insightful understanding on the waveforms. Our results show that the filter-based waveform can achieve much lower OOB compared to the standard OFDM with neglected performance degradation. Therefore, as a replacement of OFDM, the filter-based waveform is expected to be used for asynchronous transmission in future 5G networks.

\bibliographystyle{IEEEtran}
\bibliography{IEEEabrv,WaveformBib}

\end{document}